# Accessing the transport properties of pristine few-layer black phosphorus by van der Waals passivation in inert atmosphere

*Rostislav A. Doganov[1,2,3], Eoin C.T. O'Farrell[1,2], Steven P. Koenig[1,2], Yuting Yeo[1,2], Angelo Ziletti[4], Alexandra Carvalho[1], David K. Campbell[5], David F. Coker[4,], Kenji Watanabe[6], Takashi Taniguchi[6], Antonio H. Castro Neto[1,2,3], and Barbaros Özyilmaz[1,2,3,*]*

[1]Graphene Research Centre, National University of Singapore, 6 Science Drive 2, 117546 Singapore

[2]Department of Physics, National University of Singapore, 2 Science Drive 3, 117542 Singapore

[3] Graduate School for Integrative Sciences and Engineering (NGS), National University of Singapore, 28 Medical Drive, 117456 Singapore

[4]Department of Chemistry, Boston University, 590 Commonwealth Avenue, Boston Massachusetts 02215, USA

[5]Department of Physics, Boston University, 590 Commonwealth Avenue, Boston Massachusetts 02215, USA

[6] National Institute for Materials Science, 1-1 Namiki, Tsukuba, 305-0044 Japan

ABSTRACT. Ultrathin black phosphorus, or phosphorene, is the second known elementary two-dimensional material that can be exfoliated from a bulk van der Waals crystal. Unlike graphene it is a semiconductor with a sizeable band gap and its excellent electronic properties make it attractive for applications in transistor, logic, and optoelectronic devices. However, it is also the first widely investigated two dimensional electronic material to undergo degradation upon exposure to ambient air. Therefore a passivation method is required to study the intrinsic material properties, understand how oxidation affects the physical transport properties and to enable future application of phosphorene. Here we demonstrate that atomically thin graphene and hexagonal boron nitride crystals can be used for passivation of ultrathin black phosphorus. We report that few-layer pristine black phosphorus channels passivated in an inert gas environment, without any prior exposure to air, exhibit greatly improved n-type charge transport resulting in symmetric electron and hole trans-conductance characteristics. We attribute these results to the formation of oxygen acceptor states in air-exposed samples which drastically perturb the band structure in comparison to the pristine passivated black phosphorus.

Two-dimensional (2D) van der Waals crystals are expected to have huge impact on future technologies[1]. Intrinsically 2D materials have the potential to bypass major hurdles for the semiconductor industry, such as the problem of fabricating and integrating ultrathin high-quality semiconductor channels[2], or achieving high-mobility transparent and flexible thin-film transistors[3]. Until recently the only semiconductor materials that were known to remain stable down to a few layers were the semiconducting 2D transition metal dichalcogenides (TMDC)[4], e.g. $MoS_2$ and $WS_2$. The recent reports on exfoliated ultrathin black phosphorus (bP) –or phosphorene- have introduced a new material to the family of 2D semiconductors and have aroused great excitement[5–8].

Phosphorene is the second known 2D crystal besides graphene which is formed by a single chemical element and can be exfoliated from a bulk van der Waals crystal.[4] The monolayer has an optical band gap of 1.2 eV which decreases exponentially with increasing number of layers to about 0.3 eV in the bulk.[9] Ultrathin bP has been shown to exhibit excellent semiconducting electronic properties, in particular a high hole mobility of about 100 $cm^2V^{-1}s^{-1}$ for the thinnest samples and up to 1000 $cm^2V^{-1}s^{-1}$ for 10 nm thick channels, with an on-off ratio in the range $10^3$-$10^5$ at room temperature.[5–7] Unlike the semiconducting TMDCs the band gap in bP is direct for all number of layers which makes the material particularly promising for optoelectronic applications.[10] The anisotropic in-plane transport and the superconducting phase in the bulk at high pressures[11–13] could offer further exciting directions for phosphorene research. However, initial studies have shown that exfoliated bP crystals are not fully stable at ambient conditions.[6,7,14] A considerable deterioration of the surface roughness one hour after exfoliation has been observed using atomic force microscopy (AFM).[6] The effect also becomes apparent under optical microscopy if the exfoliated crystals are left in air for 24 hours.[14] Recent theoretical studies show that oxygen defects can be easily introduced under normal environmental conditions.[15] Currently the disintegration of few-layer bP limits the possibility to study monolayer crystals[7] and also raises the question of how surface deterioration effects impact transport behavior. A method for passivation of ultrathin bP crystals without prior exposure to ambient air and without chemical precursor treatment is necessary for further studies and practical applications of phosphorene.

In this article we report that dry transfer of both graphene and hexagonal boron nitride (hBN) onto ultrathin bP exfoliated in an inert Ar gas environment preserves the pristine bP crystals and limits the degradation upon exposure to ambient air and fabrication chemicals. Using AFM and Raman spectroscopy we show that the passivated bP crystals are preserved under ambient conditions. We fabricated bP field effect transistors (FETs) that allow direct comparison of passivated and exposed regions of the ultrathin crystal. This allowed us to access the transport properties of the pristine material and to study in-situ how oxidation affects the transport properties. Passivated bP channels show a hugely (10 to 100-fold) improved electron mobility at room temperature resulting in symmetric electron and hole trans-conductance characteristics. Ambipolar behavior is consistent with theoretical band-structure calculations of pristine black phosphorus.[12,16] By in-situ comparison we show that the dominant p-type behaviour and suppressed electron transport of earlier reports is due to the formation of oxygen acceptor states upon exposure of the exfoliated crystals to air. In-situ comparison of the same bP crystal allows us to exclude the effect of sample-to-sample variation and bulk crystal quality.

**Results**

We start by cleaving and exfoliating ultrathin bP crystals onto $SiO_2$/Si wafers using micromechanical exfoliation in an Ar-filled glovebox with an $O_2$ and $H_2O$ concentration of less than 2 ppm. Crystals of thickness 4-10 nm are identified under an optical microscope in the glovebox. For the passivation of ultrathin bP we use the dry transfer method developed for the fabrication of high quality graphene heterostructures[17–19]. In short, either graphene or hBN is exfoliated onto $SiO_2$/Si wafers coated with polymethylglutarimide (PMGI) and polymethylmethacrylate (PMMA), which serve as release and support polymer layers respectively. After identifying a suitable graphene or few-layer hBN crystal the PMGI is developed and the passivation crystal held by the PMMA polymer layer is put on a transfer slide and brought into the glovebox. To complete the transfer the graphene or hBN is lowered onto the exfoliated bP using a micromanipulator in the glovebox and the support PMMA layer is removed. After the transfer we observe the formation of interface bubbles (see Supplementary Information) which indicates a clean 2D interface between the bP and the encapsulating crystal[19,20].

To demonstrate that the transfer of graphene or few-layer hBN preserves the pristine ultrathin bP crystals we perform AFM measurements and Raman spectroscopy of adjacent exposed and protected surfaces after removing the sample from the inert gas environment. Figures 1a and 1b show AFM scans of an exfoliated ~5nm thick bP crystal covered with graphene and exposed to ambient air for 10 min and 24 hours respectively. From the acquired AFM images it is evident that the unprotected surface of the ultrathin bP crystal develops significant roughness while parts of the bP that are under graphene do not show any noticeable surface change. In Figure 1c similar behavior is observed for bP samples covered with hBN. In Figures 1d and 1f we compare the average surface roughness ($R_a$) versus time for the passivated and unprotected regions. $R_a$ is calculated as the arithmetic average of the absolute values of the surface height deviations, $R_a = \sum_i^N |\Delta z_i|/N$, where $\Delta z$ is surface height deviation measured from the mean surface plane and N is the total number of data points in the scan. In the following we concentrate the discussion on the sample passivated with monolayer graphene because we expect the measured $R_a$ value to be closer to that of the underlying bP crystal. In Figure 1d we observe that in the first scan, 10 min after exposure to air, for the exposed surface $R_a= 0.43$ nm, compared to only 0.29 nm for the passivated. Over a period of 8 hours the surface covered with graphene does not show any significant change while the $R_a$ value for the exposed area further increases to above 2 nm. In Figure 1e we show the height histograms of the two studied regions. The protected part of the crystal under the graphene has a Gaussian height distribution, which preserves its form over time. Only a small shift of the average height with ~0.5 nm is observed for the passivated region which could be due to surface deterioration at the bottom bP surface at the $SiO_2$ substrate. On the other hand, the exposed region has a skewed height distribution even after 10 min, which over times broadens and develops tails for heights above 8 nm, indicating the proliferation of peaks on the bP surface. Upon further exposure the roughness on the unprotected region congregates into thick droplet-like structures, which become observable under optical microscopy (Figure 1c and Figures 2a,b). For the graphene covered regions after 48 hours in ambient conditions degradation can be observed near the edges of the passivating crystal (Figure 2b).

The quality of the passivated bP is further confirmed by Raman spectroscopy 48 hours after exposure to air. As can be seen from the optical image in Figure 2a at this point the exposed part of the bP is fully degraded. The spectrum in Figure 2d verifies that the passivated region has clear $A_g^1$, $B_{2g}$, and $A_g^2$ peaks as expected for bP[6,7]. Figure 2c shows a spatial Raman map of the normalized intensity of the bP $A_g^1$ peak. It is evident that the bP peak is only present under the graphene passivation. The fully degraded part of the sample does not show any noticeable Raman signature for bP. Similar results are obtained when comparing bP samples passivated with hBN.

We now compare the charge transport characteristics of the passivated and exposed ultrathin bP. To fabricate devices for electrical transport measurements as a passivating layer we use few-layer insulating hBN. Contacts are deposited on both sides of the hBN strip to measure the pristine part of the bP crystal. Additional contacts are made to probe the exposed region of the same crystal for direct comparison of the electronic transport in each region. The device structure and an optical image of a typical sample are shown in Figures 3a and 3b.

Figure 3c shows four-terminal conductance against back gate voltage ($V_g$) at room temperature of both encapsulated and exposed regions for two devices of different thickness (sample #1 is 4.5 nm and sample #2 is 5.7 nm). For the exposed channels on the hole conduction side, at negative $V_g$=-70 V, the sheet conductance increases to >2 µS while at positive $V_g$, on the electron side, the conductance remains an order of magnitude smaller. The observation of a higher hole conductance and suppressed electron transport for the exposed region of the device is consistent with previously reported bP FETs and demonstrates dominant p-type behavior[5–8,10]. In comparison, the passivated part of the device has sheet conductance of >5 µS at $V_g$ = 70 V thereby exhibiting approximately symmetric electron and hole trans-conductance centered around zero $V_g$. In the passivated channel the conductance on the electron side improves by a factor of more than 10, while the current in the off state and the p-type conductance remain largely unchanged except for a negative shift in the relative doping level by $\Delta V_g$ ~ 30 V. The shift of the threshold voltage in the exposed channels on the hole conduction side indicates a p-doping of the ultrathin black phosphorus upon exposure to ambient air.

From the four-terminal conductance in Figure 3 we estimate the field effect mobility using

$$\mu_{FE} = \frac{L}{W} \frac{1}{C_g} \frac{dG}{dV_g} \quad . \tag{1}$$

Here $C_g$ is the backgate capacitance; G is the electrical conductance; W and L are respectively the length and width of the channel. The $SiO_2$ back gate does not allow us to fully reach the conduction bands and linear transport regime ($dG/dV_g$=const) so the actual mobility of our devices at room temperature is expected to be higher than what we estimate here (see Supplementary Information). On the hole side of our devices the mobility is in the region $\mu_{FE} >$ 10 $cm^2V^{-1}s^{-1}$ at room temperature for both the passivated and exposed channel. At T = 200 K we obtain hole mobility of 118 $cm^2V^{-1}s^{-1}$ for the exposed channel and 86 $cm^2V^{-1}s^{-1}$ for the passivated channel. These values are consistent with what has been previously reported for a bP FETs of similar thickness[5,7]. The slightly higher hole mobility extracted in the exposed region is due to the fact that the channel is more p-doped allowing us to reach further into the valance band. In the passivated channel at positive $V_g$ the improved electron conductance leads to significantly improved electron mobility. At room temperature we obtain $\mu_{FE} >$ 10 $cm^2V^{-1}s^{-1}$ on the electron side while the values in the exposed region remain 10 to 100 times lower. At T=200 K in the passivated channel we obtain an electron mobility of 62 $cm^2V^{-1}s^{-1}$, while the values in the exposed region do not exceed 5 $cm^2V^{-1}s^{-1}$. An enhancement of electron transport for the passivated channels has been reproduced in all measured samples and the direct comparison with the exposed region clearly demonstrates the effect of the passivation.

In the passivated bP channels at room temperature we observe hysteresis in the conductance depending on the $V_g$ sweep direction (Figure S5) which is similar to that in the exposed regions and previous reports[6]. This suggests that the hysteresis is largely caused by charge traps at the $bP/SiO_2$ interface rather than the degradation of the top surface. We now turn to Figure 4 and the I-V characteristic of sample #2 from Fig. 4. In order to exclude the effect of the gate sweep hysteresis the following discussion is based on transport measurements at 200 K (see Supplementary Information). In Figure 4a we plot the two-terminal conductance versus $V_g$ for the passivated and exposed channels at source-drain voltage $V_{sd}$=50 mV. In the passivated channel we observe symmetric electron and hole transport as discussed above. The on-off ratio is ~$10^5$ and is limited by the gate leakage current. At negative $V_g$ the I-V characteristic of both the passivated and exposed channels shows slight deviations from linearity (Figure 4b), which indicates the presence of a Schottky barrier between the metal contact and the bP. On the electron conduction side the I-V characteristic of the passivated and exposed channels is markedly different. In Figure 4d a Schottky behavior is observed for the exposed channel. In contrast, on the electron side of the passivated bP, for $V_{sd}$ up to ~0.5 V, we measure a linear dependence of $I_{sd}$ with $V_{sd}$ (Figure 4c). The resistance in this regime decreases with increasing $V_g$, as expected for a FET operating as a variable resistor in linear mode. For $V_{sd}$ above ~0.5 V we see the onset of current saturation indicating a pinch off in the conduction channel. In the saturation regime the device exhibits a residual linear increase of $I_{sd}$ with $V_{sd}$, which is independent of the gate voltage. From the output conductance $g_d=\partial I_{sd}/\partial V_{sd}$ in the saturation regime (shown in the inset of Figure 4c) we determine a finite drain output resistance of ~$10^6$ Ω.

The improved electron conductance and field effect mobility, together with the presence of a linear and saturation regime in the I-V characteristic at negative $V_g$, demonstrate the formation of a clear electron conduction channel and the operation of the passivated bP device in electron accumulation. It is unlikely that the improved electron transport is due to differences in the electrical contact and Schottky barrier height because the metal electrodes for probing both the passivated and exposed channels are deposited on exposed parts of the bP crystal under the same fabrication conditions. Therefore we propose that the response of the passivated channel corresponds to the behavior of pristine bP, while the transport in the exposed channel is affected by degradation consistent with the AFM and Raman spectroscopy observations. The major hallmarks of the degradation in the exposed channel are the deterioration of transport on the electron side, moderate p-doping, and essentially unaffected transport on the hole side. These transport results suggest the formation of electron traps in the exposed bP, i.e. the formation of unoccupied defect states within the band gap, such that under positive electrostatic gating the accumulated electrons fill the defect gap states and the Fermi level cannot be shifted to the conduction band.

Electron trap states in bP were shown to be induced by oxygen point defects in a recent theoretical study of monolayer phosphorene[15]; it was found that two configurations of chemisorbed oxygen, the diagonal oxygen bridge ($O_{b-d}$) and the horizontal oxygen bridge ($O_{b-h}$) lead to the formation of electron traps. We now consider how the exposure to ambient atmosphere would produce this behavior by consideration of first principles density functional theory (DFT) calculations. In Figure 5 we show electronic band structure of mono and few-layer bP with a chemisorbed oxygen diagonal bridge $O_{b-d}$ defect. In monolayer the oxygen atom introduces two states within the bandgap which have O-$p_x$ and O-$p_z$ orbital character. With increasing number of layers, as the gap decreases, both the $p_x$ and $p_z$ states plunge and hybridize with the bP valence band (Figures 5b,c). Since the $p_z$ state is filled it does not affect the conduction of electrons, whereas the $p_x$ state is an electron trap, consistent with experimental observation. We note that semilocal DFT calculations are known to underestimate the bandgap and the calculated electron trap state is highly dispersive which may be an artifact of the finite size of the calculation supercell. Nevertheless, despite these shortcomings the calculation qualitatively reproduces the inclusion of oxygen defects to pristine multilayer bP and suggests that the experimental observations are primarily the result of the formation of the oxygen bridge defects on the exposed channel, which are absent in the passivated channel.

**Discussion**

In summary, we have shown that the dry transfer of graphene and hBN can be used to passivate ultrathin bP prior to any exposure to ambient atmosphere. By AFM and Raman spectroscopy we have shown that the passivation preserves the pristine bP surface which normally degrades in ambient air. Ultrathin bP devices which are exfoliated in an inert Ar gas environment and are passivated with hBN prior to any exposure to air exhibit greatly enhanced electron transport, resulting in ambipolar charge transport  The fact that in the passivated samples the hole mobility is 30% higher than the electron mobility is in good agreement with the 20 to 30% higher effective electron mass $(m_x m_y)^{1/2}$ in the plane [12,16]. We suggest that the low electron conduction in non-passivated bP devices and the resulting p-type behaviour, which has so far remained unexplained, could be due to electron trap states which arise with the degradation of the surface upon exposure to air.

The transport results presented here make pristine bP the only 2D semiconductor crystal besides $WSe_2$ and $MoSe_2$ for which symmetric electron and hole transport has been demonstrated using a $SiO_2$ backgate and same contact metal for both electron and hole injection[21,22]. For example, symmetric trans-conductance characteristics in $MoS_2$ and most of the other TMDCs has so far been obtained only by ionic liquid gating[23], a PMMA substrate[24], or different contact metals for hole and electron accumulation[21]. The relatively high electron and hole mobilities, the large on-off ratio, and the symmetric ambipolar behavior could enable the fabrication of CMOS-like digital logic elements using encapsulated phosphorene.

We highlight that the method for passivating exfoliated bP presented here can be used to study the pristine properties of other 2D crystals which are unstable in air, such as metallic TMDCs and layered semiconductors like GaSe and $Bi_2Se_3$[4]. Graphene and hBN are known to be excellent gas barriers[25,26] and their ultrathin character allows minimum interference with the properties of the underlying crystal. Furthermore the dry transfer in an inert gas environment does not require any chemical precursors or exposure to water which is often the case for atomic layer deposition or other thin film deposition techniques. Here we have performed a comparative study of only partly passivated samples, but for future applications our passivation method needs to be further developed to allow electrical contact to fully passivated devices.

# Methods

## Sample preparation

Black phosphorus (bP) crystals were purchased from smart-elements GmbH and exfoliated onto <100> Si wafers with a 300 nm thermal oxide in an Ar-filled glovebox with an $O_2$ and $H_2O$ concentration of less than 2 ppm. No degradation of the bP crystals is observed in the Ar gas environment (see Supplementary Information). Thin bP flakes were identified using an optical microscope in the glovebox. The hBN crystals used for the encapsulation were exfoliated from bulk hBN single crystals grown by a Ba-BN solvent method[27]. The dry transfer was performed in the glovebox by the process described in the main text. For device fabrication a layer of PMMA is spin coated onto the bP/BN heterostructure in the glovebox and the sample is then removed from the Ar environment for e-beam lithography and thermal evaporation of contacts (5 nm Ti and 80 nm Au).

## AFM and Raman measurements

AFM scans were acquired using a Bruker Dimension FastScan microscope in tapping mode. Between the scans the bP samples were kept in a class 1000 cleanroom with controlled 50% relative humidity. Raman spectroscopy was performed in ambient conditions in the backscattering configuration with a 532 nm laser excitation.

## Electrical Transport Measurements

Electrical transport measurements were performed either in DC, or using AC lockin amplifiers together with a DL instruments 1211 current preamplifier at low frequency (<3 Hz). All electrical transport measurements were performed in vacuum.

## Density-functional theory calculations

Black phosphorus multi-layers were modeled within the framework of density-functional theory, as implemented in the siesta package.[28,29] The generalized gradient approximation of Perdew, Burke and Ernzerhof is used for the exchange-correlation functional.[30] The electronic core is accounted for by using ab-initio norm-conserving pseudopotentials with the Troullier-Martins parameterization[31] in the Kleinman-Bylander form[32]. The charge density was assumed to be independent on spin. The basis sets for the Kohn-Sham states are linear combinations of numerical atomic orbitals (double zeta polarised basis).[33,34] The charge density is projected on a real-space grid with an equivalent cutoff energy of 250 Ry to calculate the exchange-correlation and Hartree potentials. A Monkhorst-Pack[35] scheme with 4x4x1 points is used to sample the Brillouin Zone.

ACKNOWLEDGEMENTS

This work was supported by the Singapore National Research Foundation Fellowship award (RF2008-07-R-144-000-245-281), the NRF-CRP award (R-144-000-295-281), and the Singapore Millennium Foundation-NUS Research Horizons award (R-144-001-271-592; R-144-001-271-646). AHCN acknowledges the NRF-CRP award "Novel 2D materials with tailored properties: beyond graphene" (R-144-000-295-281). The calculations were performed at the GRC computing facilities. A.Z. and D.F.C. acknowledge NSF grant CHE-1301157.

**Author contributions**

R.A.D. and S.P.K. initiated the research, fabricated the samples, performed the AFM and Raman measurements, and obtained initial transport results. E.C.T.O.F. performed the detailed electrical transport measurements and transport analysis presented in the article. Y.Y. helped with the transfer process. A.Z., A.C., D.K.C., and D.F.C. performed the DFT study. K.W. and T.T. grew the hBN crystals used for the encapsulation. A.H.C.N and B.O. directed the research project. All authors contributed to the preparation of the manuscript.

**Competing financial interests**

The authors declare no competing financial interests.

FIGURES

**Figure 1 AFM study of the surface integrity of passivated and exposed ultrathin bP in ambient conditions**. **(a-b)** AFM scans of a 5 nm thin bP crystal partly covered with graphene. The images are acquired respectively 10 min and 24 hours after exposure to ambient conditions. The white dashed lines outline the passivating graphene crystal. **(c)** AFM scan of a 10 nm thin bP crystal partly covered with hBN after 5 days in ambient conditions. The scale bar in all images is 4µm. **(d)** Average roughness ($R_a$) versus time in ambient conditions for the passivated (blue triangles) and exposed (red squares) bP surface of the bP / graphene sample from (a-b). The dashed lines are a guide to the eye. **(e)** Height distribution of the exposed (top) and encapsulated bP surface (bottom) at different times after exposure to ambient conditions for the bP / graphene sample from (a-b). **(f)** Average roughness ($R_a$) versus time for the passivated (blue triangles) and exposed (red squares) bP surface of the gP / hBN sample from (c). The dashed lines are a guide to the eye.

**Figure 2 Raman spectroscopy characterization of passivated and exposed ultrathin bP 48 hours after exposure to ambient conditions.** **(a)** Optical image of the bP / graphene sample from Fig.1 after 48 hours in ambient conditions. **(b)** AFM image of the bP/graphene sample after 48 hours in ambient conditions. The scale bar is 4µm in both images. **(c)** Spatial Raman map of

the normalized intensity of the bP $A_g^1$ peak. **(d)** Raman spectrum of the passivated region of the crystal showing the bP $A_g^1$, $B_{2g}$, and $A_g^2$ peaks, as well as the graphene G and 2D peaks. The inset shows a close up on the bP peaks.

**Figure 3. Comparison of encapsulated and exposed ultrathin bP FETs. (a)** A schematic three-dimensional illustration of the device geometry. **(b)** Optical image of a typical device with outlines of the ultrathin bP crystal (black dashed-dotted line) and the passivating hBN crystal (blue dashed line). The scale bar is 3μm. **(c)** Four-terminal conductance versus back gate voltage ($V_g$) of the passivated and exposed channels of two bP FETs at $V_{sd}$ = 50 mV and T=300 K. Sample #1 thickness ~4.5 nm and sample #2 thickness ~5.7 nm. For the exposed region of sample #1 and #2 the backgate voltage has been shifted by -10 V and -40 V respectively. Field effect mobilities were extracted from the line fit (black dashed line).

**Figure 4 I-V characteristic of the passivated and exposed channels of a bP FET. (a)** Two-terminal sheet conductance versus back gate voltage ($V_g$) of sample #2 from Fig.3 for the passivated (solid blue curve) and exposed (solid red curve) channel at T=200 K. **(b)** I-V characteristic of the device on the hole side for the passivated (solid blue curve) and exposed (solid red curve) channel taken at $V_g$ = -60 V and $V_g$=-50 V indicated by a vertical black line in (a). **(c)** Source-drain current ($I_{sd}$) versus source-drain voltage ($V_{sd}$) for the passivated channel on the electron conduction side at $V_g$ from +30 V to +75 V in steps of 5 V (solid curves). The inset shows the output conductace $\partial I_{sd}/\partial V_{sd}$ at $V_g$ = +75 V, $V_g$ = +65 V, and Vg= +55 V. **(d)** $I_{sd}$ versus $V_{sd}$ for the exposed region on the electron conduction side at $V_g$ from +30 V to +75 V in steps of 5 V.

**Figure 5 Electronic band structure of oxidized few-layer black phosphorus.** Band structure of ultrathin black phosphorus with a chemisorbed diagonal oxygen bridge defect ($O_{b-d}$) from monolayer (1L) up to three layers (3L). Occupied band are represented by solid black lines and unoccupied by solid red lines. The bands of the pristine black phosphorus are overlaid as shade, and the labels indicate the p character of the oxygen related bands at the G point.

# Figure 1

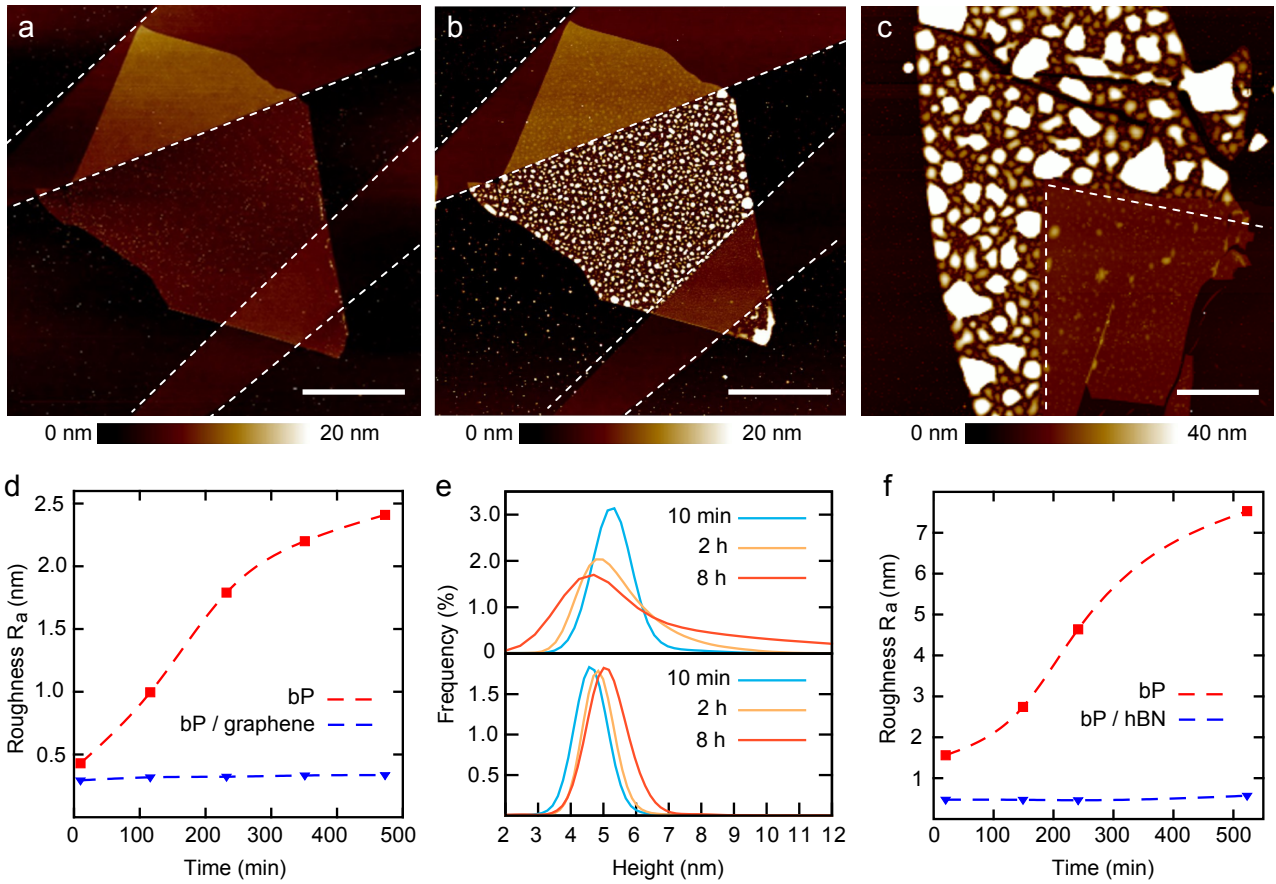

Figure 2

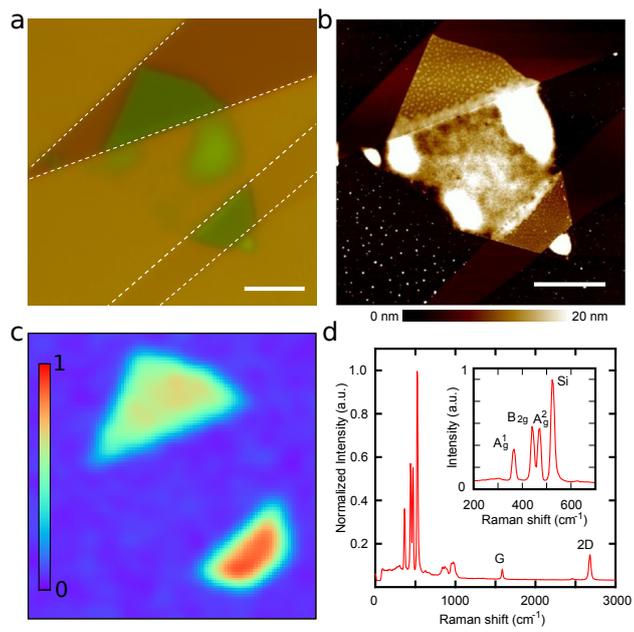

Figure 3

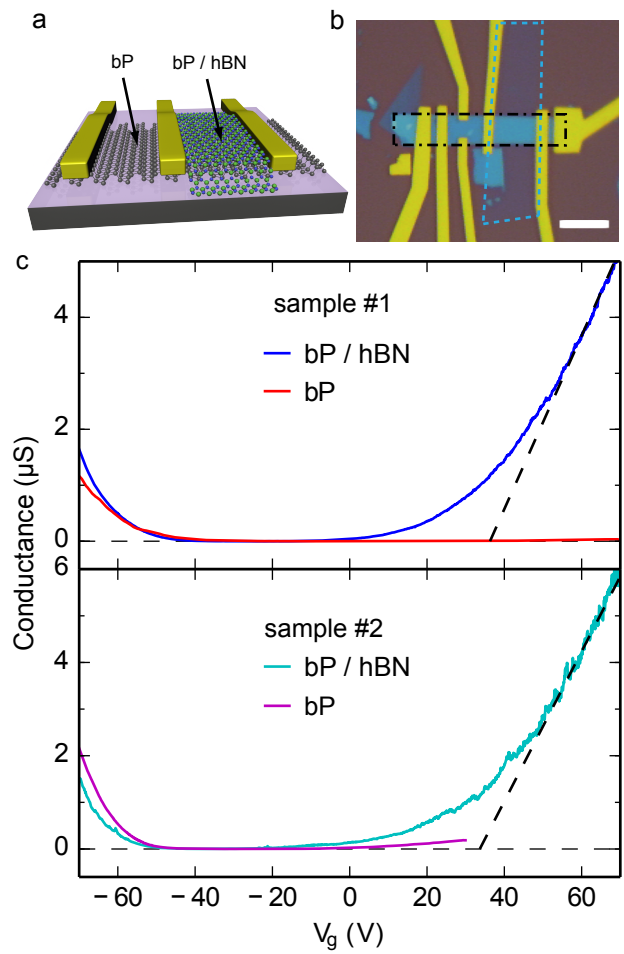

Figure 4

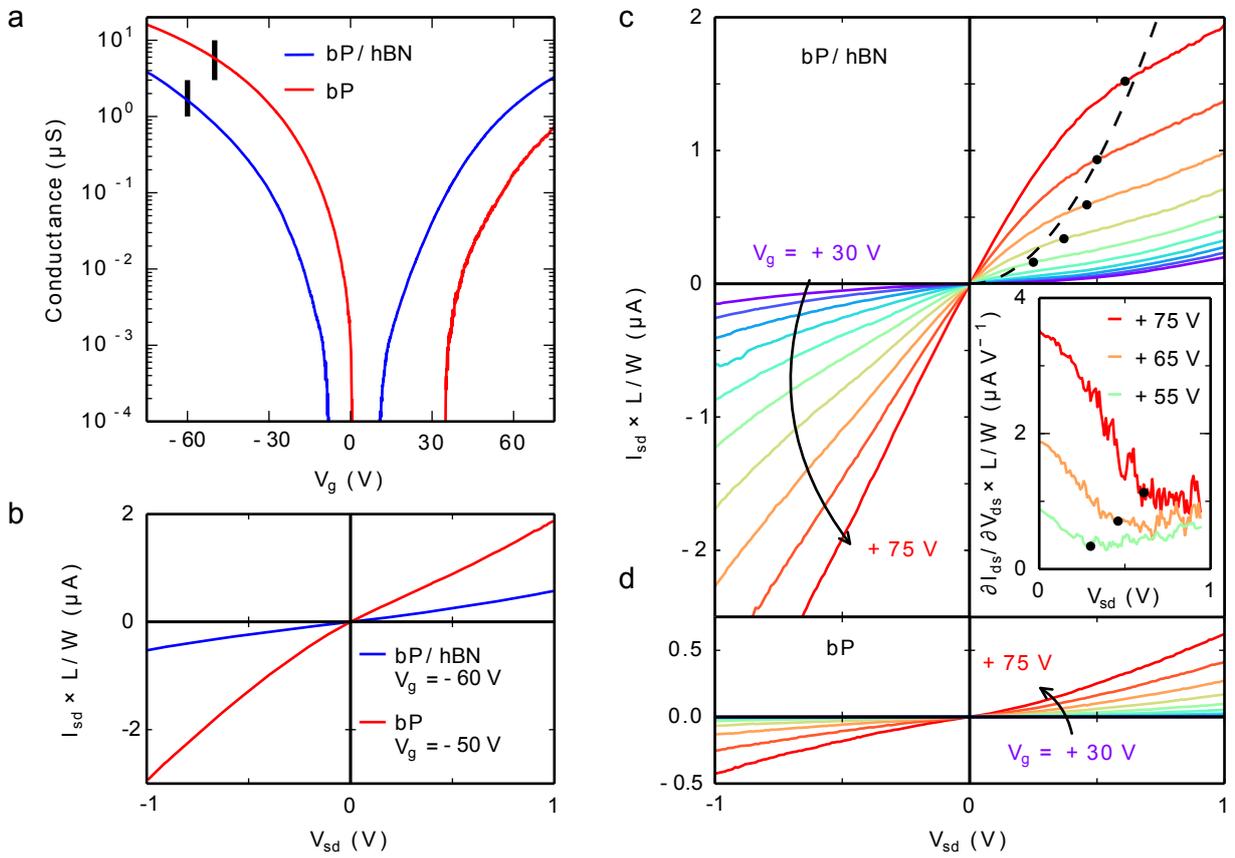

Figure 5

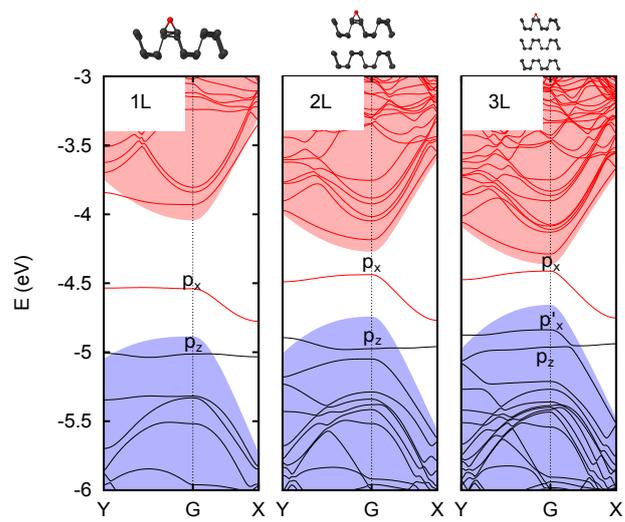

**Supplementary Information for:**

**Accessing the transport properties of pristine few-layer black phosphorus by van der Waals passivation in inert atmosphere**

*Rostislav A. Doganov*[1,2,3]*, Eoin C.T. O'Farrell*[1,2]*, Steven P. Koenig*[1,2]*, Yuting Yeo*[1,2]*, Angelo Ziletti*[4]*, Alexandra Carvalho*[1]*, David K. Campbell*[5]*, David F. Coker*[4,]*, Kenji Watanabe*[6]*, Takashi Taniguchi*[6]*, Antonio H. Castro Neto*[1,2,3]*, and Barbaros Özyilmaz*[1,2,3,*]

[1]Graphene Research Centre, National University of Singapore, 6 Science Drive 2, 117546 Singapore

[2]Department of Physics, National University of Singapore, 2 Science Drive 3, 117542 Singapore

[3] Graduate School for Integrative Sciences and Engineering (NGS), National University of Singapore, 28 Medical Drive, 117456 Singapore

[4]Department of Chemistry, Boston University, 590 Commonwealth Avenue, Boston Massachusetts 02215, USA

[5]Department of Physics, Boston University, 590 Commonwealth Avenue, Boston Massachusetts 02215, USA

[6] National Institute for Materials Science, 1-1 Namiki, Tsukuba, 305-0044 Japan

**Supporting Information**

1. **Absence of bP degradation in Ar gas**

   We do not observe any degradation when exfoliated black phosphrosus (bP) crystals are kept in an Ar-filled glovebox with $H_2O$ and $O_2$ concentration of less than 2 ppm. This is consistent with studies showing the importance of both $H_2O$ and $O_2$ in the degradation process[1]. In Figure S1 we show optical images of a thin crystal taken 18 days after exfoliation which do not show any noticeable change in the surface integrity.

2. **Formation of transfer bubbles at the graphene / bP interface**

   The transfer of graphene and hBN onto bP leads to the formation of transfer bubbles which indicate a self-cleaning effect[2]. Upon heat annealing the small transfer bubbles congregate into larger ones as has been reported for transfer of graphene onto other two dimensional crystals[2,3]. In Figure S2 we show AFM scans of graphene on bP before and after heat annealing, demonstrating the formation of the transfer bubbles.

3. **Hysteresis in charge transport measurements**

   Transport measurements of exfoliated bP using a $SiO_2$ back gate show significant hysteresis which depends on the back gate sweep rate, the sweep direction, and the starting or extreme point of the sweep. Hysteresis is significantly suppressed by measuring at lower temperature.[4] Figure S3 shows the effect of hysteresis on the drain current modulation at $T = 280$ K and 200 K.

4. **Electric field effect mobility estimation**

The 300 nm SiO$_2$ gate does not allow us to fully reach the conduction or valance band and reach the linear conduction regime. In Figure S4 we plot the electric field effect mobility versus gate voltage for sample #2 from the main text. From the figure it can be seen that even at the highest gate voltage the mobility does not saturate and is still increasing.

Supplementary Information References

**Supplementary Figures Cpations**

**Figure S1 (a)** Optical image of an ultrathin bP crystal taken in the glovebox ~30 min after exfoliation. **(b)** Optical image of the same ultrathin crystal after ~18 days in the glovebox without any exposure to ambient air. The scale bar is 10 µm.

**Figure S2** AFM image of an ultrathin bP crystal covered with monolayer graphene showing the formation of transfer bubbles. **(a)** AFM image of the sample as transferred. **(b)** The same sample after 4 hour of Ar+$H_2$ (9:1) gas annealing at 300 C.

**Figure S3** Source-drain current ($I_{sd}$) versus gate voltage ($V_g$) for the FET device discussed in Fig. 4 of the main text at T=280 K (green curves) and T=200 K (blue curves) for both sweep directions of the back gate voltage: from positive to negative gate voltage (solid curves) and from negative to positive (dashed curves).

**Figure S4** Field effect mobility (as defined by Eq. (1) in the main text) versus gate voltage ($V_g$) of sample #2 from the main text for the passivated (solid blue line) and exposed (solid red line) channel.

# Figure S1

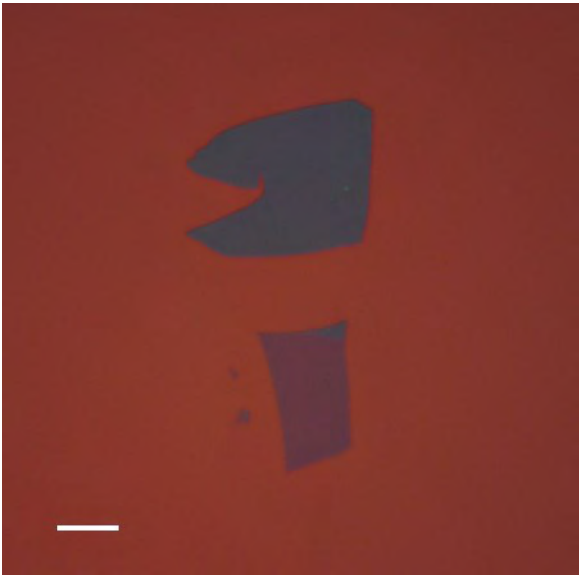 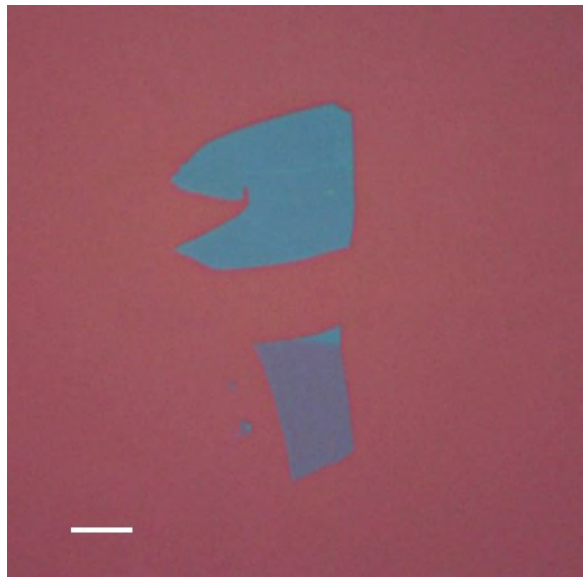

# Figure S2

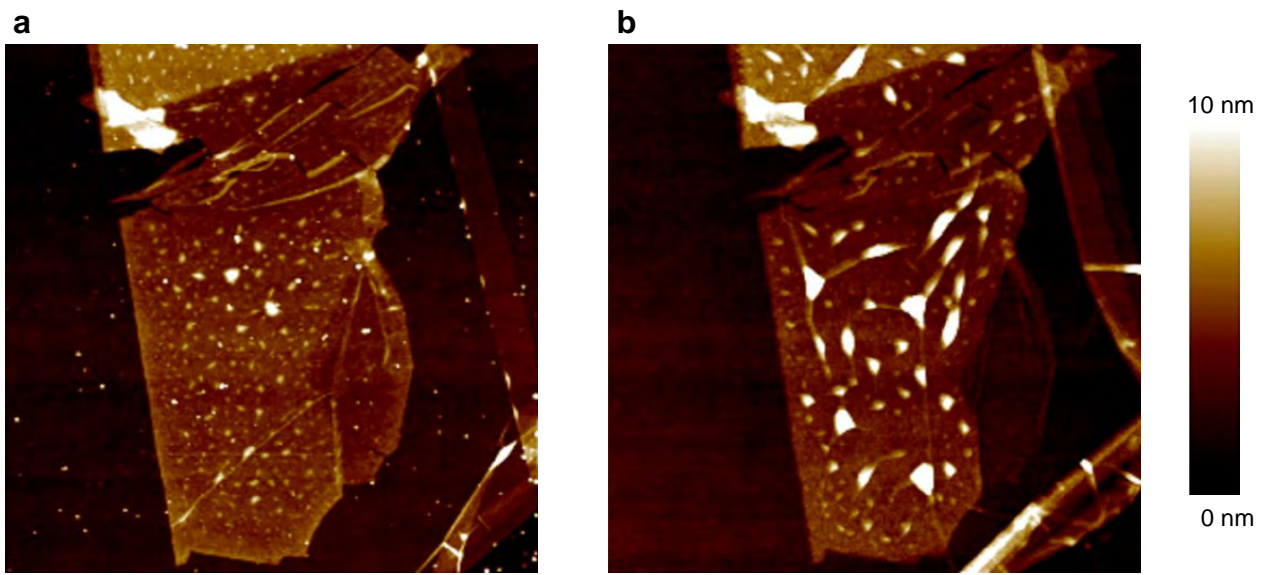

# Figure S3

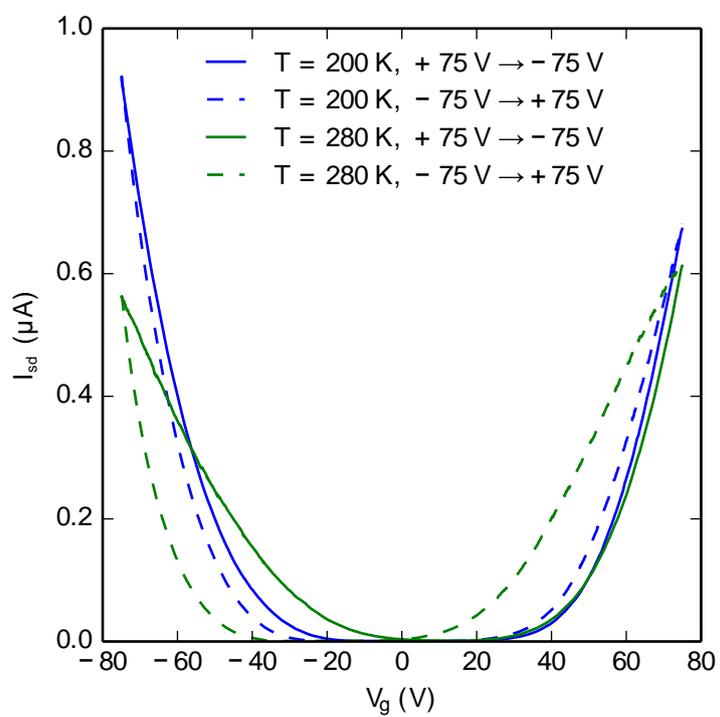

# Figure S4

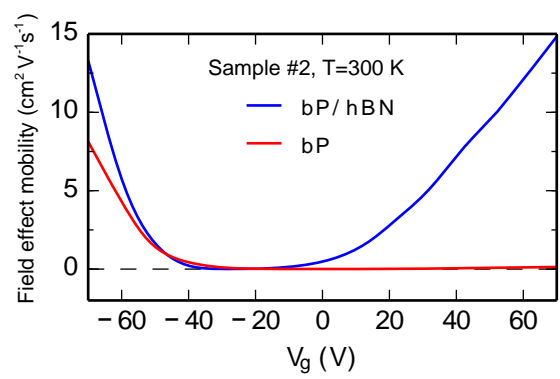